\documentclass[12pt,preprint]{aastex}

\slugcomment{Submitted to the Astrophysical Journal Letters}

\shorttitle{A Novel Method for Measuring the EBL}
\shortauthors{Georganopoulos, Sambruna, Kazanas, Cillis, Cheung, Perlman, Blundell, Davis}

\begin{document}

\title{A novel method for measuring the  extragalactic background light:   Fermi application to the lobes of  Fornax A}

\author{ Markos Georganopoulos\altaffilmark{1,2}, Rita M. Sambruna\altaffilmark{2}, Demosthenes Kazanas\altaffilmark{2}, Analia N. Cillis\altaffilmark{2},  Chi C. Cheung\altaffilmark{2}, Eric S. Perlman\altaffilmark{3}, Katherine M. Blundell\altaffilmark{4}, David S. Davis\altaffilmark{1,2}}

\altaffiltext{1}{Department of Physics, Joint Center for Astrophysics, University of Maryland Baltimore County, 1000 Hilltop Circle, Baltimore, MD 21250, USA}
\altaffiltext{2}{Astrophysics Science Division, NASA Goddard Space Flight Center, Code 660, Greenbelt, MD 20771, USA}
\altaffiltext{3}{Department of Physics and Space Sciences, Florida Institute of Technology, 150 West University Boulevard, Melbourne, FL 32901, USA}
\altaffiltext{4}{Department of Physics, University of Oxford, Keble Road, Oxford OX1 3RH, UK}

\begin{abstract}
We describe  a new method for measuring the extragalactic background light (EBL)
 through the detection of $\gamma$-ray  inverse Compton (IC) emission due to  scattering of the EBL photons  off relativistic electrons in the  lobes of  radio galaxies.
 Our method has no free physical parameters
 and is a powerful tool when the lobes are  characterized by a  high energy sharp break or cutoff in their  electron energy distribution (EED). We show that such a feature will produce  a high energy  IC  `imprint'  of the  EBL spectrum in which the  radio lobes are embedded, and show how this  imprint  can be used to derive the EBL.  We apply our method to  the bright nearby radio galaxy Fornax A, for which we demonstrate, using WMAP and EGRET observations,  that the EED of its lobes is characterized by a  conveniently located cutoff, bringing the IC EBL emission into  the {\sl Fermi} energy range.   
 We show that {\sl Fermi} will set upper limits to the optical EBL and measure the more elusive infrared EBL.
 \end{abstract}


\section{Introduction \label{section:intro}}

The EBL  that permeates the Universe in the optical-IR
 is closely connected to the history of structure formation. 
It  is composed  by the cosmic infrared background (CIB)  peaking at  $\lambda\sim 100 \mu m$ ( dust-reprocessed starlight) and the cosmic optical background  (COB)  peaking at $\sim 1 \mu m$ (starlight), and its  level  is still unknown to a factor of a few,
due to the dominance of foregrounds
 such as the interplanetary dust emission  and Galactic  emission
    \citep{hauser01,kashlinsky05}. Robust lower limits   to the EBL level  have been set by galaxy counts (e.g. Dole et al. 2006). 
 Useful upper limits in the 1-10 $\mu$m range  can be obtained by
   using the TeV blazar emission  to set limits on the EBL by modeling its attenuation due to pair production with EBL photons  (e.g. Stecker, de Jager, Salamon 1992).
   This method assumes that the intrinsic blazar TeV spectrum cannot be arbitrarily hard (e.g. Aharonian et al. 2006). 
 However, this can  be circumvented  \citep{katarzynski06, stecker07, aharonian08}, significantly relaxing  the EBL limits derived
 (Mazin \& Raue 2007).

 We propose  to measure the EBL  by detecting the high energy emission produced  when EBL photons IC-scatter off the relativistic electrons of extragalactic sources.  IC scattering of the cosmic microwave background (CMB) has  already been detected in the X-rays  from the lobes of a handful of radio galaxies (e.g. Erlund et al. 2008). In fact,  the first confirmation for the presence of the  CMB  in an extragalactic location  came from the ROSAT X-ray detection of  
  the lobes of Fornax A by Feigelson et al. (1995).
 The idea we propose  is based on the fact that, although the EBL energy density is $\sim$  only a few percent that of the CMB, 
 the CIB and COB  photons  have  energies  correspondingly  $\sim$ 10 and 1000 times  higher than those of the  CMB.
  For a source with a given EED, 
   its electrons  will  IC-scatter  the CMB and EBL photons, and the IC
emission will  have a spectral shape related  to that  of the CMB and EBL, shifted in 
frequency by $\gamma_{\rm max}^2$, where $\gamma_{\rm max}$ is the maximum Lorentz factor of the EED.  
 %
Therefore, the IC emission from  such a source  will consist of a powerful component due to CMB seed photons and two weaker  components with a power of
  a few percent that of the  CMB one, but shifted in energy by a factor $\sim 10$ and $1000$.
The cleanest spectral separation of these  IC components results from a power-law EED characterized by a  high energy cutoff.  

 If this IC  emission can be detected, we will observe the high energy  imprint  of the CMB+EBL at the location of our  source.  This imprint  will provide us with  the level and spectral shape of the  CIB+COB  at the source.
Here we show that the lobes of  radio galaxies, if
 characterized by  an EED  with  a suitably located high energy cutoff, can be used to measure the EBL at their location  through {\sl Fermi} observations, and we demonstrate our method on the  radio lobes of the nearby bright radio galaxy Fornax A.


\section{EBL $\gamma$-ray imprints: the case  of Fornax A \label{section:fornaxa}}


We exclude  blazars and radio quasars,  whose GeV emission is dominated by their pc-scale  jet  IC emission. 
The favored sources are the  extended  lobes of radio galaxies with weak  active galactic nuclei (AGN)  that are not expected to be significant GeV emitters.  Moreover weak AGN will not contribute  optical-IR seed photons to the radio lobes, beyond those contributed  by the host galaxy.
In addition to the lobe synchrotron radio emission, the  lobe IC emission off the CMB must  also be  detected,  to determine the lobe magnetic field  and the EED normalization, {\sl without resorting to the equipartition assumption. This parameter-free  determination of the EED is necessary for a one-to-one
mapping of the EBL to its IC emission.}
The only cases  for which  IC off the CMB has been  unambiguously 
identified is the  X-ray emission of the lobes of a handful of radio galaxies. Given that in these sources  the X-ray spectral index $\alpha_x<1$, the peak of the IC emission off the CMB is at higher energies, and  the peak of the IC emission of the CIB and COB
is at energies $\sim 10$ and 1000 times  higher correspondingly. 
The peak energy of  the three IC  components 
$\propto \gamma_{\rm max}^2$.  This means  that the only energy range at which
we could expect  to detect the IC emission off the CIB and COB is the GeV energy range covered by  {\sl Fermi}, and this for a relatively narrow range of $\gamma_{\rm max}\sim 10^5$ that keeps the IC  off the CMB below  the $100$ MeV lower energy threshold of {\sl Fermi}. Larger values would dominate the {\sl Fermi} band by 
 IC off the CMB, and  smaller values  would shift part or all of   the IC off the EBL below the 
$100$  MeV threshold.
Sources that satisfy the $\gamma_{\rm max}$  constraint  are  required to have no other $\gamma$-ray source within  {\sl Fermi}'s  angular resolution ($\sim 30'$ at 1 GeV). A  high Galactic latitude is desirable, because it reduces  contamination from  Galactic sources. 
Finally,  high frequency radio observations and existing EGRET limits of the $\gamma$-ray flux are needed to constraint the critical quantity   $\gamma_{\rm max}$.

\begin{figure}
\epsscale{0.9}
\plotone{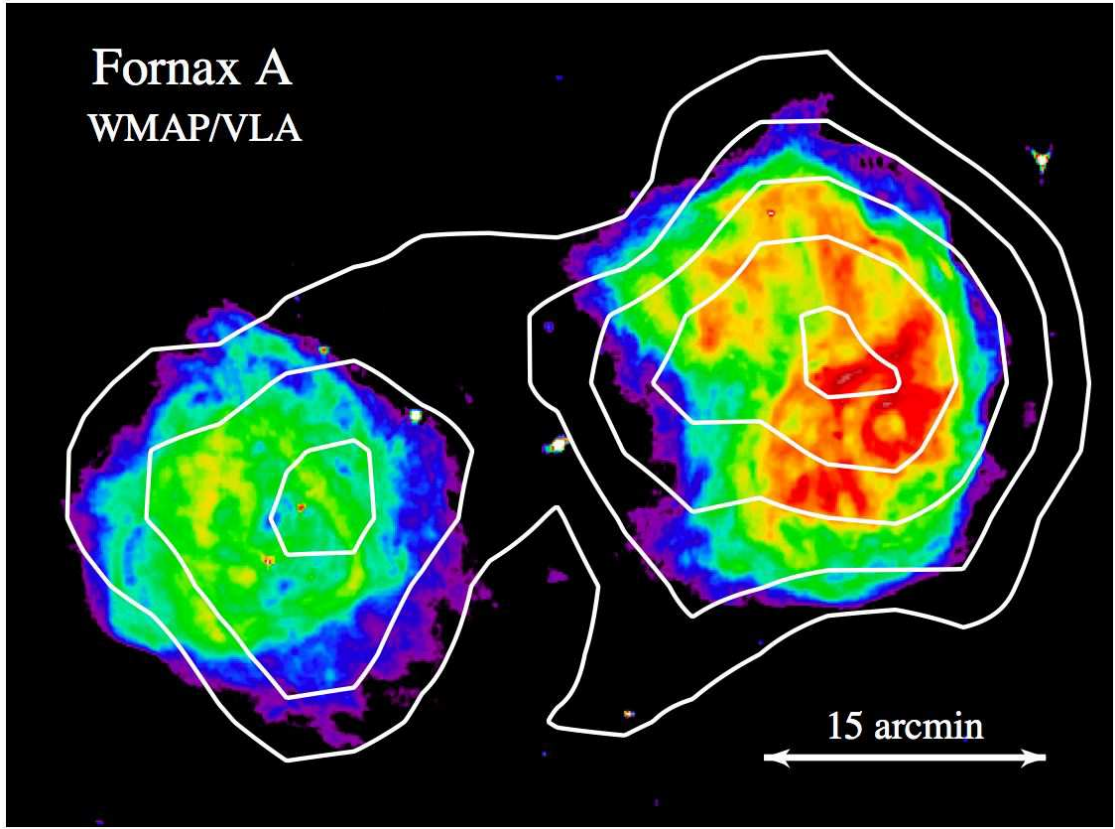}
\caption {The 1.5 GHz VLA (color; Fomalont et al. 1989) and 61 GHz WMAP (contours; Hinshaw et al. 2007)  images of Fornax A.}
\label{fig:VLAWMAP}
\end{figure}

Fornax A is a high Galactic latitude ($-57^{\circ}$) bright  radio galaxy at a distance of 18.6  Mpc 
(5.41 kpc arcmin$^{-1}$), hosted by the  massive elliptical NGC 1316 that shows a  LINER core \citep{isobe06}. 
In the radio it shows  two relaxed lobes  of  $\sim 20'$  diameter separated by  $\gtrsim 30'$ (Fig. \ref{fig:VLAWMAP}). 
 We use radio data  collected 
by Isobe et al. (2006), replacing the
extrapolated 100 MHz data point from Finlay \& Jones (1973) with an 86
MHz measurement (Mills et al. 1960).
We also use the integrated 3-yr WMAP fluxes  (Hinshaw et al. 2007) from
23 to 61 GHz, which are larger than the 1-yr
WMAP fluxes quoted  by Cheung (2007).  
The total radio spectrum has a spectral index of $\alpha_{\rm r}=0.68 \pm 0.1$ (Isobe et al. 2006), increasing to $\sim$0.8 in the WMAP  band (Hinshaw et
al. 2007). The western lobe is $\sim$1.9 times brighter than the eastern in the VLA 1.5 GHz map, but this ratio is only $\sim$1.3 in the WMAP 41 GHz and 61 GHz maps. 
The lobes have been detected in X-rays  by ROSAT \citep{feigelson95}  and XMM (only the eastern lobe was observed) with  $\alpha_X=0.62_{-0.15}^{+0.24}$ \citep{isobe06},
consistent with the radio index $\alpha_r$ (here we assume $\alpha_r=\alpha_X=\alpha=0.65$). 
The  lobe X-ray emission is  due to IC-scattered  CMB photons (the synchrotron-self Compton level is much below  the observed flux). 
Because the CMB  energy density is given, the X-ray flux  and spectral index  uniquely define the EED slope and normalization. With this at  hand, the magnetic field required to produce the radio emission is uniquely determined to $B=1.7\; \mu$G,  close to the equipartition field of $B=1.55 \mu$G (Isobe et al. 2006).

 
EGRET  provides a $2.2 \; \sigma$ detection at the $10^{-8}$ photons cm$^{-2}$ s$^{-1}$ level, for photons with energy E$>100$ MeV \citep{cillis04},  which we treat as an upper limit for the total  lobe flux.  A cutoff
is required in the lobe EED, otherwise the IC of the CMB emission would violate the EGRET upper limit.
This constrains $\gamma_{\rm max}\lesssim (3\nu_{EGRET}/4\nu_{CMB})^{1/2}\sim3 \times 10^5$, with the additional constraint that $\gamma_{\rm max}$ must be sufficiently high to extend the synchrotron emission to the WMAP energies, $\gamma_{\rm max} \gtrsim (\nu_{WMAP} / 2\times 10^6 B)^{1/2}\sim 10^5$.
A  fit of the combined constraints  requires $\gamma_{\rm max}=1.6\times10^5$,  close to the optimal $\gamma_{\rm max}\sim  10^{5}$.
 The  total lobe spectral energy distribution (SED)  and the model SED  is shown in Fig. \ref{fig:lowhigh}, where the dot-dash line is  IC emission off the CMB. Note that while none of the synchrotron and IC peaks are  detected, their location is well constrained from the WMAP and EGRET data.


The electrons in the  lobes will  also experience the
 directional photon field of the host galaxy.
The unknown source orientation affects the  IC level  through the angle of IC scattering (e.g. Dermer, Schlickeiser, \& Mastichiadis 1992) 
and by determining the distance of the lobes from the host galaxy. The combined effect results in a $\theta-$dependence of the total IC flux from both lobes  due to galactic seed photons
$f_{IC, gal}\propto [(1+\cos\theta)^2+(1-\cos\theta)^2] \sin\theta^2$, that has a flat maximum at $\theta=\pi/2$, remaining within $ 10\,\% $ of the maximum for  $ \theta\gtrsim 60^{\circ}$.
We  assume that the source axis forms an angle of  $\theta= 60^{\circ}$ 
to the line of sight, using an average   lobe projected distance of $15'$ from the galaxy.  Given the flat $\theta$-dependence of $ f_{IC, gal}$  for $\theta\gtrsim 60^\circ$, 
  our choice represents  the maximum  IC emission of the lobes due to host galaxy seed photons. 
The SED  of the host galaxy \citep{dale07}
 is comprised of two components, the first one  peaking at $\sim 1\mu$m, and the second one, $\sim 30$ times weaker,  peaking at $\sim 100 \mu $m. We approximate its SED
 as the sum of two black bodies and we plot in Fig. \ref{fig:seeds} with a dotted line the resulting isotropic equivalent  photon field intensity at the lobes.
  
For the EBL we adopt  a lower  and  an upper level (short and long dashed lines respectively in Fig. \ref{fig:seeds}) of the expected background, by roughly following  the range considered by  \cite{mazin07}.  This is meant to represent a plausible range of the yet unknown EBL.  Without loss of generality,
we choose to represent the EBL as a histogram for  reasons  that  will become apparent in \S \ref{EBL_to_imprint}.  We note here that  the lower limit is rather well established, because it relies  on galaxy counts (e.g. Dole et al. 2006).
 As can be seen from Fig. \ref{fig:seeds}, the lobe photon field at the $\lambda \lesssim 10\mu {\rm m}$ has comparable contributions from the galaxy and the COB for all plausible COB levels, while  at $\lambda \gtrsim 10\mu {\rm m}$ and up to $\sim 300 \mu {\rm m}$, the photon field is dominated by the CIB by more than a factor of ten, even for the lower EBL limit. 

\section{The $\gamma$-ray imprint  of the EBL \label{EBL_to_imprint}}

\begin{figure}
\epsscale{0.9}
\plotone{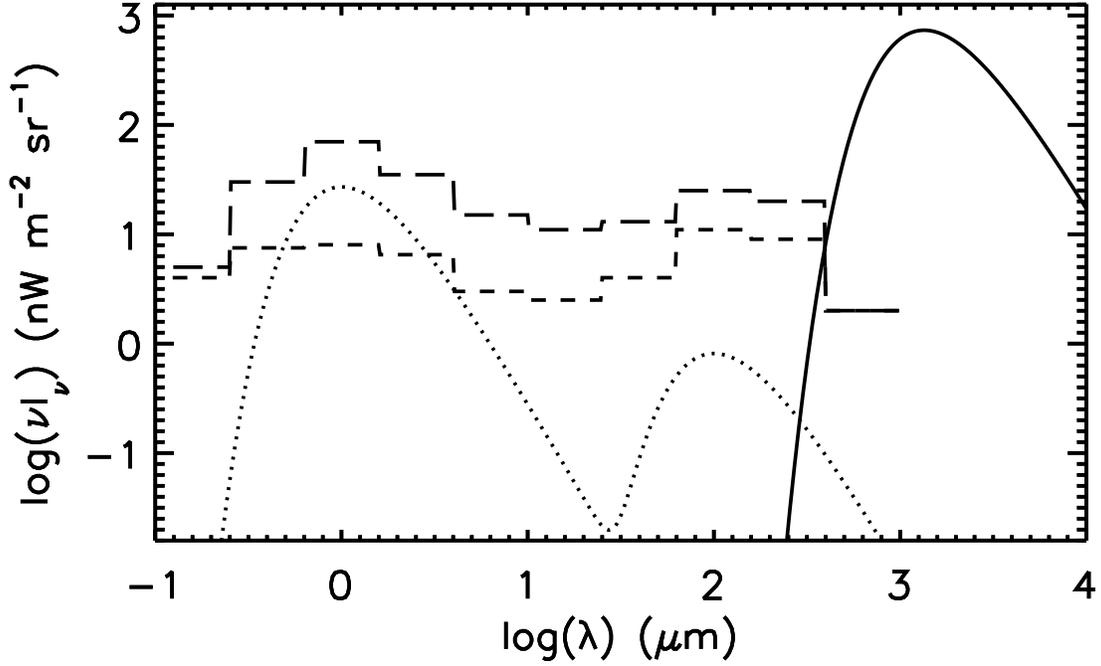}
\caption {The seed photons at the lobes of Fornax A. CMB: solid line. Host Galaxy: dotted line. Short dashed line: Low estimate of the EBL. Long-dashed line: High estimate of the EBL. 
The  $1 \mu$m and $100 \mu$m peaking components due to the host galaxy have a total intensity of 
 36.7 nW m$^{-2}$ sr$^{-1}$ and 1.1 nW m$^{-2}$ sr$^{-1}$ respectively.
 In the low EBL  case, the COB   (liberally extending to  $\lambda=10\, \mu$m ) has an integrated  intensity of 26.7  nW m$^{-2}$ sr$^{-1}$ and the CIB ($10<\lambda<10^3\, \mu$m )  an integrated intensity of 26.2 nW m$^{-2}$ sr$^{-1}$. 
In the high EBL case  the COB has an integrated   intensity of 142.7 nW m$^{-2}$ sr$^{-1}$ and the CIB an integrated   intensity of 72.8 nW m$^{-2}$ sr$^{-1}$. }
\label{fig:seeds}
\end{figure}

\begin{figure}
\epsscale{0.9}
\plotone{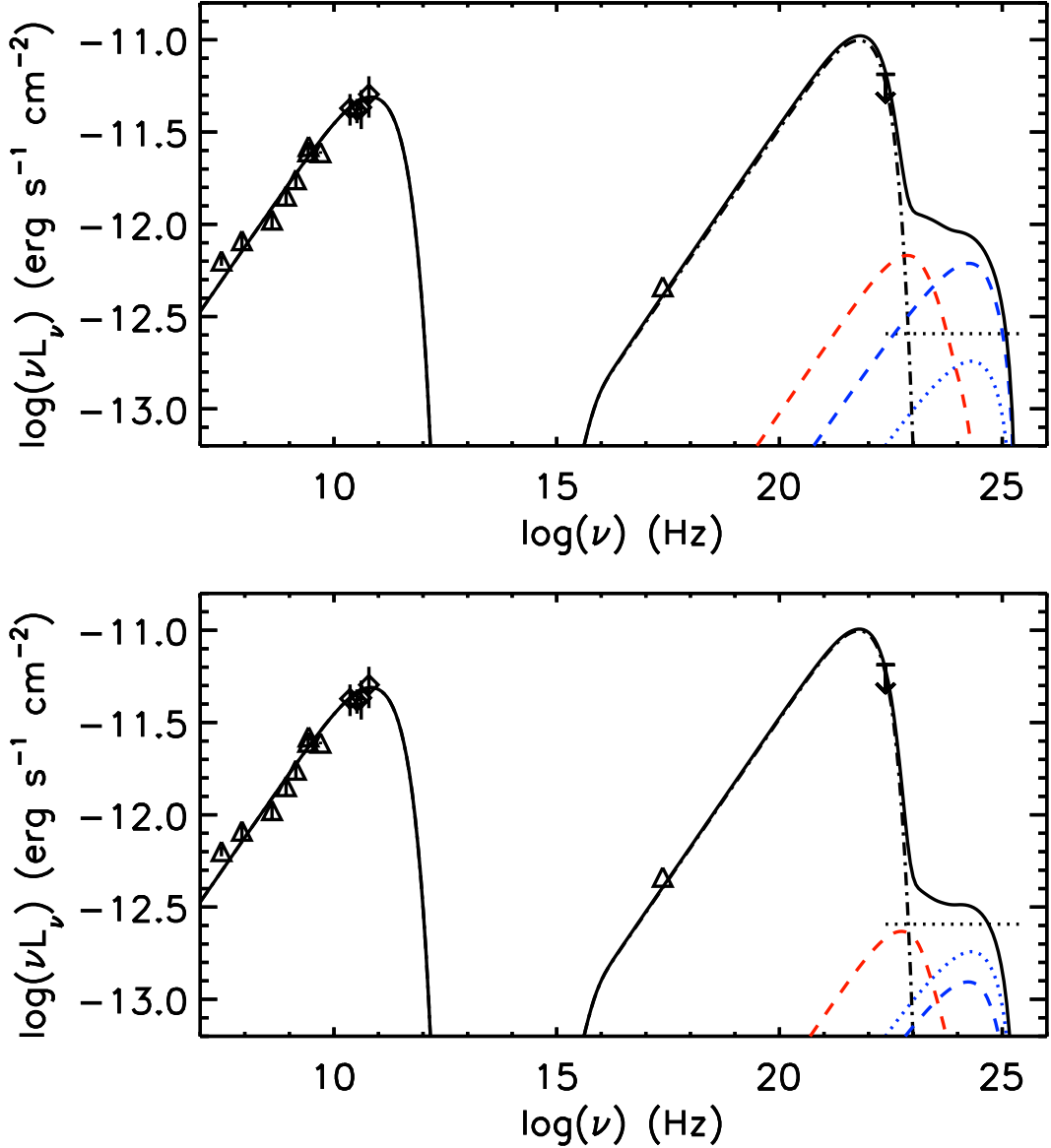}
\caption {
The radio, WMAP, and  X-ray flux from both radio lobes, as well as the EGRET upper limit. The solid line is  the model SED,
resulting from a magnetic field of $B=1.7 \; \mu$G, and a  power law  EED with slope $p=2.3 $ and maximum Lorentz factor $\gamma_{\rm max}=1.6\times 10^{5}$. As discussed in \S 
\ref{section:fornaxa}  these parameters are  strongly constrained by the data, and have a very small range in which they can vary. 
We also plot the  IC due to the CMB (dot-dash line), the CIB and COB (red and blue broken line), as well as the maximum expected level of the IC emission due to the optical photons of the host galaxy (dotted blue  line).  The black dotted line marks the 2 year, 5$\sigma$ {\sl Fermi} sensitivity limit. The lower (upper)  panel corresponds to the low (high)  level EBL.}
\label{fig:lowhigh}
\end{figure}

We present now the  SED of the IC emission  for the lower and higher EBL  cases (lower and upper panels of Fig. \ref{fig:lowhigh}).  
The  black solid line is the total SED.  The dotted blue line is the IC  emission due to the host
galaxy optical photons. The IC  emission due to the host galaxy IR  photons is too weak to appear in the plots. 
 To identify the contribution of the COB and CIB  seed photons,  
 we plot with broken blue and red lines the IC emission due to seed photons with $\lambda<10 \mu$m (five short-$\lambda$  EBL bins  in Fig. \ref{fig:seeds}) and  with $\lambda>10 \mu$m (five  long-$\lambda$  EBL bins  in Fig. \ref{fig:seeds}) respectively.
 We also plot the 2-year {\sl Fermi} 5-$\sigma$ sensitivity limit (dotted black line). 
 We see that the IC of the EBL plus galaxy seed photons is detectable in both cases, although in the lower EBL case it is only somewhat above the 5-$\sigma$  limit.

  We note that, while the IC emission due to $\lambda>10 \mu$m seed photons is CIB dominated,  the COB and galaxy IC contributions from the
 $\lambda<10 \mu$m  seed photons are comparable for both  low and high EBL cases.
 The IC emission from these seed photons dominates the flux  at $\nu\gtrsim  10^{24} $ Hz (corresponding to $\sim 5 $ GeV), with no contribution from lower energy seed photons.
 Given that the  $\lambda<10 \mu$m  seed photons are an unspecified mixture of   photons from the COB and the host galaxy, a measurement of the IC emission at $\gtrsim  5 $ GeV energies will provide us only with an upper limit for the COB level.
 The total emission at energies  $\sim 1-$few  GeV is  due to comparable contributions  of  $\lambda<10 \mu$m  and $\lambda>10 \mu$m   seed photons. The key in disentangling the contribution of seed photons of different energies is to consider that
 as the seed photon energy increases, their IC radiation reaches higher energies:  
   at a  {\sl Fermi} energy  $\epsilon_\gamma$,
  only seed photons with energy   $\gtrsim \epsilon_\gamma/  \gamma_{\rm max}^2$ 
   contribute. 
This can be used to reconstruct  the seed photon SED
  starting from the optical, needed to model the high energy part of 
  {\sl Fermi} observations, and gradually incorporating  lower energy IR seed photons at appropriately chosen levels, to model  the emission at gradually lower {\sl Fermi} energies. 



\begin{figure}
\epsscale{0.495}
\plotone{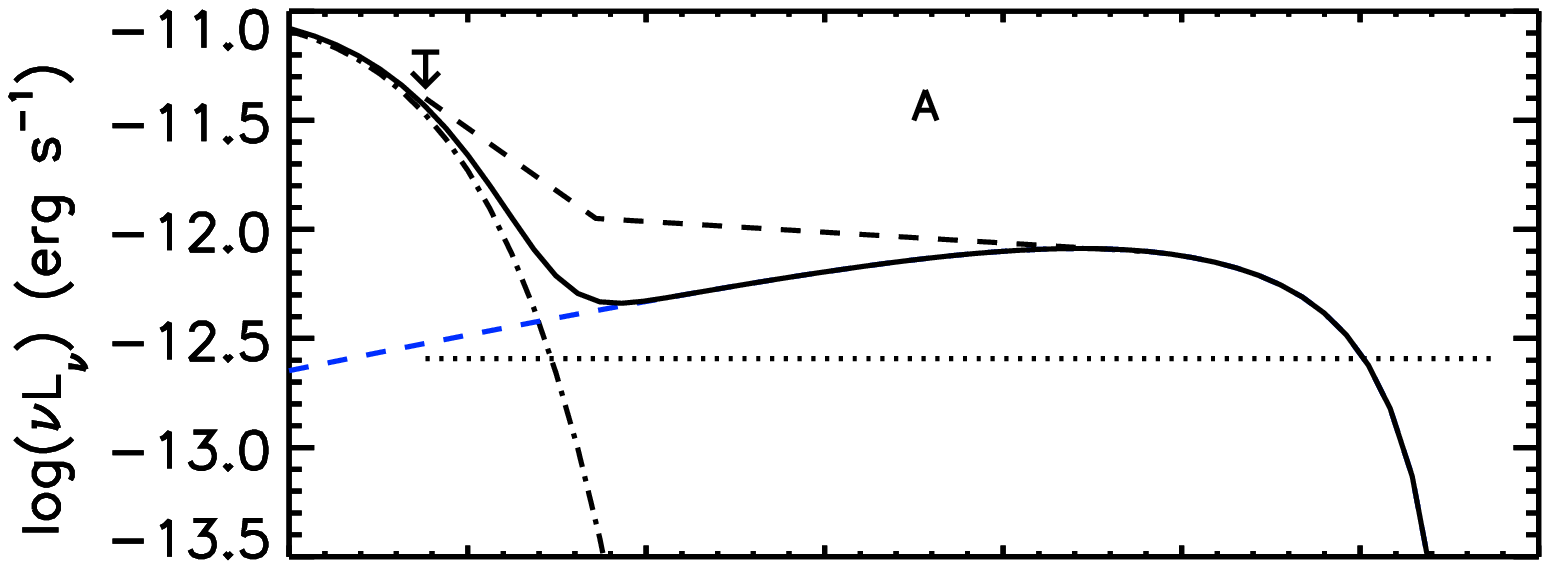}
\plotone{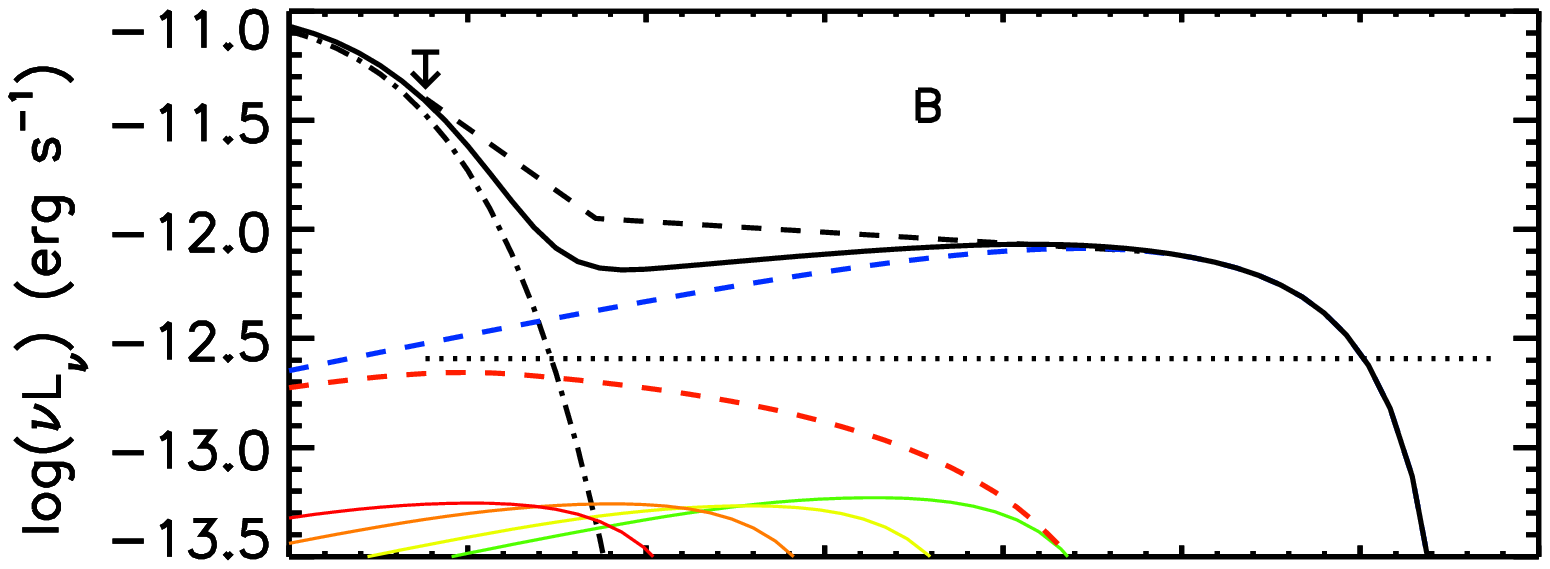}
\plotone{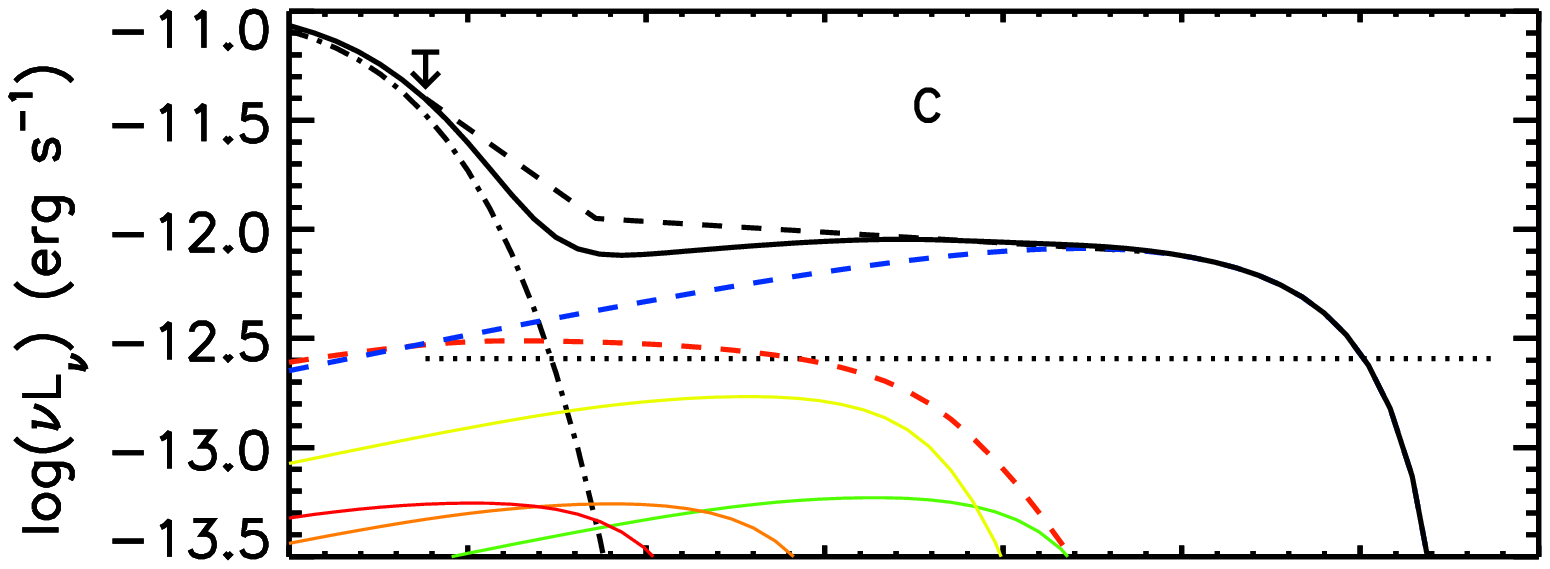}
\plotone{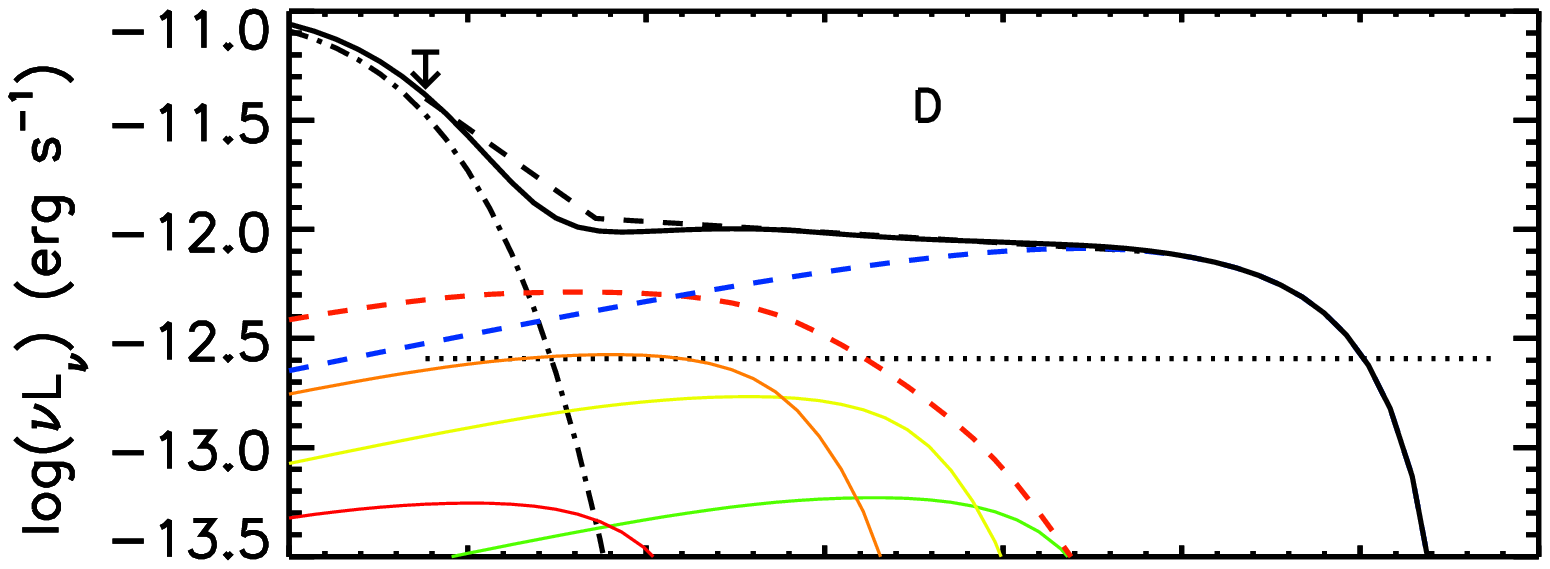}
\plotone{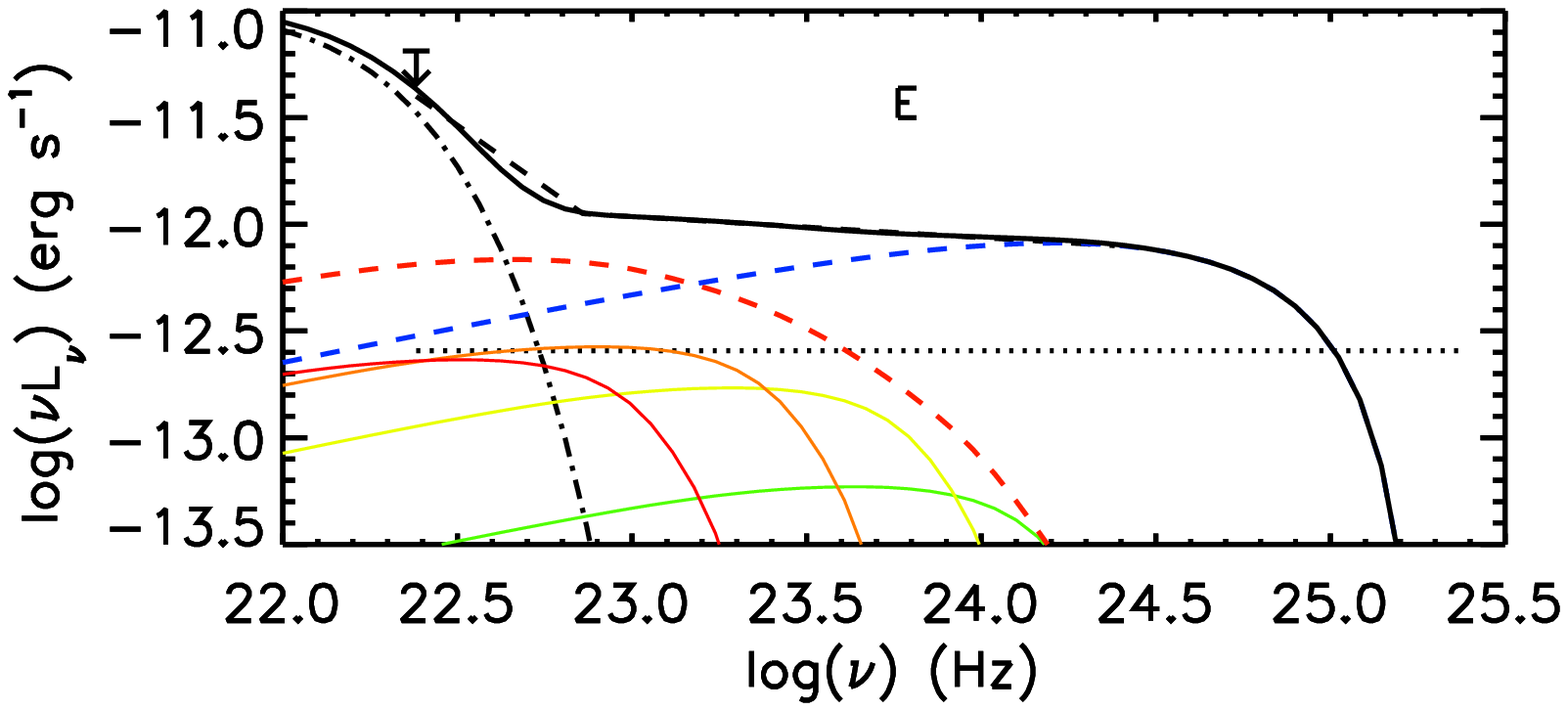}
\plotone{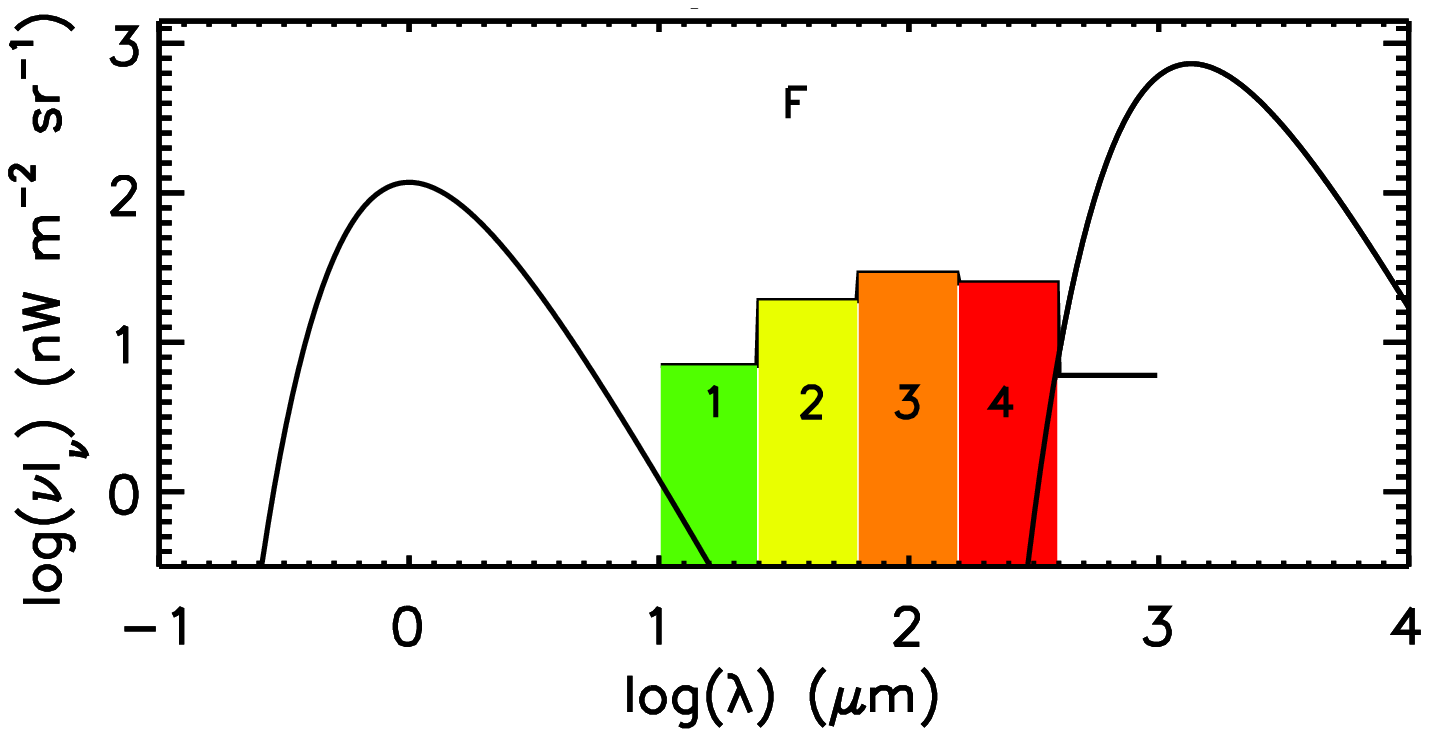}
\caption {How to go from an observed {\sl Fermi} SED to the CIB (see \S \ref {EBL_to_imprint}).}
\label{fig:inverse}
\end{figure}

  We demonstrate now how to recover the  EBL form its  $\gamma$-ray imprint, by breaking the seed photon SED into components of different energy.
  We anticipate that {\sl Fermi} will detect  a steep low energy tail due to IC-scattered  CMB photons, followed by a high energy hard component due to IC-scattered  EBL  photons. 
 Let us assume that  the EBL is at its  maximum  level (the   procedure  we  describe also applies  to lower EBL levels). This will produce an IC emission
 at the level shown in the  upper panel of Fig. \ref{fig:lowhigh}, and 
 {\sl Fermi} modeling will produce a broken power law (dashed line in Fig. \ref{fig:inverse}), soft at low  and hard at high energies.    
  As we discussed in \S \ref{EBL_to_imprint},  the $\sim$ few GeV emission is due to  an unspecified mixture of host galaxy optical seed photons and the COB, and modeling it can only provide us with an upper limit for the COB
  total intensity.
We assume for simplicity that the sum of these has  a blackbody shape peaking
at $\lambda=1\mu$m, and we adjust  its amplitude to match the $\sim 5-10$ GeV ($\sim 10^{24}-10^{24.5}$ Hz)  {\sl Fermi} level.  This blackbody intensity is plotted as a solid line at Fig. \ref{fig:inverse}F and  has a total intensity of $159.9$ nW m$^{-2}$ sr$^{-1}$. The resulting IC GeV emission is plotted in  Fig. \ref{fig:inverse}A.
Note  that our initial COB  seed photon intensity in Fig. \ref{fig:seeds} (by summing  up the five  $\lambda<10\,\mu$m bins)  is 142.7 nW m$^{-2}$ sr$^{-1}$, below our derived upper limit.

Note that the optical  blackbody  underproduces the lower energy {\sl Fermi} flux. 
This cannot be remedied by increasing its normalization, because it would  overproduce the
 $\sim 5-10$ GeV flux. Lower energy seed photons are needed.
In  Fig. \ref{fig:inverse}B  we include at a low level seed photons at the four bins with $\lambda>10\mu$m. The four thin color  lines correspond to the contributions of the same color CIB bins in Fig.  \ref{fig:inverse}F.  The broken red line is  the total emission of  these four bins,  the broken blue line is the contribution of the  $1\mu$m blackbody, and the black line is the total that has to match the observations. 
The intensity of the highest frequency seed photon bin (green) is adjusted so that the total
emission at $\gtrsim 10^{24} $ Hz, the peak energy of this component, matches the observed flux
(this produces the small difference between the broken  blue and the black line at $\gtrsim 10^{24} $ Hz). 
We continue this process in Fig.s  \ref{fig:inverse}C-\ref{fig:inverse}E  by
increasing in the intensity  of the other three energy bins, going from higher to  lower seed photon energies. The final SED of Fig.
\ref{fig:inverse}E is produced by the photon intensity shown in Fig. \ref{fig:inverse}F. 
This is the CIB photon intensity we recovered and it should be compared to the initial CIB (long dash line in Fig. \ref{fig:seeds})  that provided us with the hypothetical {\sl Fermi} observations. The initial seed photon  intensities of the COB and CIB, as well as those derived through the above procedure are given in  Table \ref{tab:EBL}. 
The recovered CIB  intensity  is  close to the initial one, although individual bins can have substantial differences.   


This toy-fitting procedure  is presented to demonstrate that the CIB (and a COB  upper limit)
 can be recovered from {\sl Fermi} observations, and to outline the  principles that the actual fitting procedure should incorporate. A more realistic  scheme would  start with two EBL components
(an optical and an IR) and adjust its amplitudes by fitting the {\sl Fermi} SED. One then would split the CIB
bin to subsequently higher number of bins, as long as increasing the number of CIB bins improves substantially  the fit of the {\sl Fermi} data.  In this way, the detail to which we  recover the CIB depends on the quality of the {\sl Fermi} data.


\section{Conclusions \label{section:discussion}} 

We have presented a novel method for measuring the EBL through   the detection of  the IC
emission resulting from the upscattering of EBL photons  by relativistic electrons in the lobes of radio galaxies. The requirements that the sources  must fulfill (extended, high Galactic latitude, X-ray detection of the IC emission due to CMB seed photons, sharp break of the EED, no significant AGN GeV emission) quickly narrow down the number of  candidate  sources.  Using existing multiwavelength data (radio, WMAP, X-ray, EGRET), we show that in the case of  Fornax A  the  EBL imprint is detectable by {\sl Fermi}, even for the lowest expected EBL level.   Our method is parameter free, in the sense that  all  physical parameters are directly derived from observations.  The radiating electron production mechanism (e.g. electron shock acceleration and/or electrons created in hadronic processes)  is not relevant for our work, since we derive the EED, regardless of it origin,  from multiwavelength data.
  {\sl Fermi} is observing in scanning mode,
so observations of Fornax A are guaranteed. Additional multiwavelength observations in the future can help refine further the derived EBL.
We also outline a procedure for solving the inverse problem of going from the upcoming  {\sl Fermi} observations to the EBL determination.
Our method will determine the CIB (and set upper limits on the COB),  finally measuring  this cosmologically important quantity.

\acknowledgements

We thank Nils Odegard for providing the WMAP images.

\clearpage
\begin{deluxetable}{ccccccc}
\tablecaption{Photon Intensity in the lobes (in nW m$^{-2}$ sr$^{-1}$)}
\tablewidth{0pt}
\tablecolumns{7}
\tablehead{
  & COB &CIB 1 
& CIB 2 & CIB 3 & CIB 4  & CIB \\
 &   &
&  &   &  & Total}
\startdata
Initial& $142.7$  & $11.0$ & $11.0$ & $27.6$ & $23.0 $
 & $72.6$ \\ 
 Recovered& $ <159.9$ & $6.4$ & $17.5$ & $26.7$ & $23.0 $
 & $73.6$\\ 
 \enddata
\label{tab:EBL}
\end{deluxetable}

\clearpage

\end{document}